*NASA / JWST Solar System Working Group Focus Group on TNO Science*

# Physical Characterization of TNOs with JWST

Alex Parker[1], Noemi Pinilla-Alonso[2], Pablo Santos-Sanz[3], John Stansberry[4], Alvaro Alvarez-Candal[5], Michele Bannister[6], Susan Benecchi[7], Jason Cook[1], Wesley Fraser[5,8], Will Grundy[9], Aurelie Guilbert[10], Bill Merline[1], Arielle Moullet[11], Michael Mueller[12], Cathy Olkin[1], Darin Ragozzine[13], Stefanie Milam[14]

**a:** aparker@boulder.swri.edu
1: Southwest Research Institute, Department of Space Studies, Boulder, CO, 80302.
2: University of Tennessee, Department of Earth and Planetary Sciences, Knoxville, TN, 37996.
3: Instituto de Astrofísica de Andalucía (IAA-CSIC), Granada.
4: Space Telescope Science Institute, Baltimore, MD.
5: Observatório Nacional - MCTI, Rio de Janeiro, Brazil.
6: University of Victoria, Victoria, BC, Canada.
7: Planetary Science Institute, Tucson, AZ 85719.
8: Queen's University Belfast, Astrophysics Research Centre, Belfast, Northern Ireland.
9: Lowell Observatory, Flagstaff, AZ 86001.
10: Institut UTINAM, Besançon, France.
11: NRAO, Charlottesville, VA 22903.
12: SRON Netherlands Institute for Space Research, Groningen, Netherlands.
13: Florida Institute of Technology, Physics and Space Sciences, Melbourne, FL, 32901.
14: NASA Goddard Space Flight Center, Greenbelt, MD.

## Abstract

Studies of the physical properties of Trans-Neptunian Objects (TNOs) are a powerful probe into the processes of planetesimal formation and solar system evolution. JWST will provide unique new capabilities for such studies. Here we outline where the capabilities of JWST open new avenues of investigation, potential valuable observations and surveys, and conclude with a discussion of community actions that may serve to enhance the eventual science return of JWST's TNO observations.

# Introduction

The distant trans-Neptunian populations are remarkably valuable tracers of the chemical and dynamical evolution of our solar system. Overlapping and mixed dynamical classes have been subjected to a variety of histories; some have clearly been subjected to significant rearrangement by interactions with the giant planets (Gomes et al. 2005, Morbidelli et al. 2005, Tsiganis et al. 2005, Levison et al. 2008, Parker 2015), while others appear to have been nearly untouched for the entirety of the age of the solar system (Parker & Kavelaars 2010, Dawson et al. 2012). Understanding the physical properties of these worlds can help constrain models of solar system cosmogony by probing the radial profiles of chemical species in the primordial disk and tracing how the planet-formation processes that grew these bodies varied across the disk.

Physically characterizing these worlds is a challenging prospect. They are typically small, and frequently have very dark surfaces. This, coupled with their extreme distance, makes them extremely faint. The population has only been studied in earnest for two decades, and the census of known bodies remains replete with biases. Because some TNO sub-populations dynamically overlap one another, assigning any given TNO to a single host population can be difficult; often, only a conditional probability of membership can be assigned. With its recent flyby of the Pluto system and near-future KBO encounters, TNO science is beginning to have an *in situ* exploration component to complement studies made from Earth environs.

The JWST will bring about a new era in physical characterization of TNOs. In particular, it will provide an unprecedented capability to investigate the diversity of surface compositions of these extremely cold and distant objects. While JWST will not be as efficient a tool to measure radiometric diameters and albedos of TNOs as it will be for warmer objects, it can perform very useful characterization of the warm *regions* on TNOs and thereby significantly improve constraints on albedos and thermal properties for objects with diameters determined by other methods. JWST's spatial resolution is not sufficient at thermal wavelengths to open many new avenues for studying the extremely common binary TNOs, but is sufficient, for example, to marginally separate Pluto and Charon. In this document, we will briefly describe both the strengths and limita-tions of JWST for TNO science, further building upon the outline presented in Norwood et al. (2015).

For any science investigation, it is reasonable to consider the relative fraction of available time required from a facility, and compare that to the historic fraction of time dedicated to similar investigations. For space telescope missions such as *Hubble*, *Spitzer*, and *Herschel*, TNO science has typically won 1-2% of available observing time to date. Given the 5-year design mission lifetime for JWST, a comparable allocation would result in roughly $2 \times 10^6$ seconds being available for TNO science. Given the roughly $10^3$ currently-known TNOs, an equal portioning results in $2 \times 10^3$ seconds per known TNO. While it may be hoped that TNO (and Solar System) science are awarded a larger fraction of time on JWST, 1000 seconds provides a useful benchmark to compare to the requirements of any proposed investigations.

## TNO Science with JWST

**Temperatures, Diameters, Albedos and Thermal Properties.**

Because the typical temperature of a TNO is of order ~50K, the thermal peak of their emission spectrum is around ~100 *µ*m. JWST's spectral coverage ends at 28.5 *µ*m; long enough to detect the blue tail of TNO thermal emission, but insufficient to bracket the thermal peak. However, the slope of TNO thermal emission in the MIRI operating range is very sensitive to temperature. If we conservatively assume that MIRI relative photometry is also uncertain at the 3% level per filter, that equates to noise-equivalent temperature differences of just 0.5K and 1K, respectively, for a 50K object observed in the MIRI 18 or 21 *µ*m and the 25.5 *µ*m filters. If the relative photometry is more precise than 3% (as it should be), then the temperature measurements possible with MIRI will be extremely accurate. Such measurements can resolve the degeneracy between albedo and beaming-parameter (or other thermal characteristics) that is intrinsic to the thermal-radiometric method (*e.g.* Lellouch *et al.* 2013), and could provide very accurate constraints on thermal inertia and roughness for objects with well-known diameters and rotation vectors (such as Haumea and some close binary systems). Some TNOs also show anomalous and rather mysterious excess short-wavelength thermal emission (Makemake, *e.g.* Lim *et al.*, 2010, and Eris, Kiss *et al.* 2015) which could be characterized in detail using MIRI.

Coupling JWST observations with archival *Spitzer* and *Herschel* observations (*e.g.* Stansberry *et al.*, 2008; Lellouch *et al.*, 2013) will allow results from those missions to be refined, and significantly so in some cases. For example, Sedna has not yet been detected at wavelengths shortward of its thermal emission peak, but plausibly could be using JWST/MIRI. Such data might reveal the presence of low-albedo regions and thereby evidence for seasonal volatile transport on Sedna. While the epochs of JWST observations and those from *Spitzer* and *Herschel* do not overlap, it is relatively straightforward to simultaneously fit multi-epoch data (as has been done in synthesizing the *Herschel* and *Spitzer* data sets, for example), subject to the assumption that seasons or other processes have not affected the intrinsic thermal properties of the body.

ALMA offers the possibility of thermal observations contemporaneous with JWST, and the combined data sets would bracket the TNO thermal peak. Roughly 500 currently-known Centaurs and TNOs can be detected with sufficient signal-to-noise ratios for accurate radiometric determination of diameters from ALMA (Moullet et al. 2011) and a commensurate survey with JWST would result in a combined data set with particular power to enhance our understanding of the sizes, albedos and thermal properties of a larger or different sample of TNOs than has already been characterized. For example, the sensitivity of ALMA and JWST/MIRI would allow the characterization of smaller TNOs, helping us understand those objects that are more similar in size to the Centaur and Jupiter-family comet populations thought to derive from the Kuiper Belt.

The proliferation of occultation networks (including dedicated telescope networks such as TNORecon[1], which is designed to observe several multi-chord TNO occultations per year at the start of full-scale operations; Marc Buie, private comm.) will likely make them competitive tools for building new statistically-useful catalogs of accurate and precise TNO diameters above and beyond what has been achieved via thermal radiometric observations and direct imaging. TNOs with highly accurate diameters from occultations will provide important benchmarks for verifying results from radiometric measurements in general, and would significantly bolster the value of JWST (and ALMA) radiometry. To the extent that data from *Spitzer*, *Herschel*, ALMA, occultations, and JWST can be obtained and synthesized for TNO binaries, it should be possible to significantly improve on the accuracy, precision, and sample-size of such systems for which we have bulk-density measurements.

---

[1] http://tnorecon.net/

### Binary Discovery and Characterization.

Lacking a critically-sampled imager at its shortest-wavelength coverage, JWST will not provide any power in excess of HST in terms of ability to resolve tightly-bound binary TNO systems. Shortward of 2 µm, NIRCam is extremely undersampled; at 2um, where the imager is moderately well sampled, the PSF FWHM is ~0.068", similar to WFC3 UVIS and lower than ACS/HRC. The stability of JWST's PSF and accuracy of its dither pointing may permit more aggressive image reconstruction methods than have been possible with HST, but this remains to be demonstrated. At mid-IR wavelengths, JWST's spatial resolution (as low as ~1" at the longest wavelengths) is insufficient to resolve all but the rarest, widest, and most collisionally sensitive binary TNO systems (Parker et al. 2011, Sheppard et al. 2012, Parker et al. 2012).

One key distinguishing feature of TNO binary systems is the common optical colors of the components (Benecchi et al. 2009). It would be valuable to use JWST's ability to perform detailed compositional characterization in the NIR to determine if TNO binary surfaces are indeed composed of identical materials. Such characterization could be carried out for systems undergoing mutual events (such as Sila-Nunam, Grundy et al. 2012), or on the sample of binary systems with mutual orbits wide enough for JWST to directly resolve both components with NIRCam; currently of order 30 systems have well-characterized mutual orbits that make them candidate targets, and ongoing surveys may identify and characterize more.

### Surface compositions.

JWST's spectral sensitivity in the IR makes it a powerful tool for measuring surface compositions of TNOs. A variety of molecular ices display spectral features in the IR, and the contrast between surfaces that are indistinguishable in the optical are often in-creased (Benecchi et al. 2011). Broadly, JWST can provide the following for TNO compositions:

1. Spectroscopy with NIRSpec will allow us to identify currently undetected volatiles on the surface of dwarf planets such as Eris, Makemake, and Pluto. Those present will be identified by their fundamental absorption bands between 3 - 5 $\mu$m. The high sensitivity of JWST and NIRSpec will permit the study of rotational light curves of these dwarf planets in the region of the fundamental bands of the volatiles. JWST may also identify irradiation products on the surface of these objects (Brown et al. 2015).

2. JWST will allow us to confirm the presence of specific volatile species on the surface of other large TNOs, such as nitrogen on Sedna (Emery et al. 2007, See Fig 1), methane or ethane on Quaoar (Shaller et al. 2007 Dalle-Ore et al. 2009), hydrated ammonia on Quaoar (Jewitt et al. 2004), and methane on Varuna (Lorenzi et al. 2014). Figure 1 illustrates model JWST NIR reflectance spectra of three hypothetical Sedna surfaces given a 10,000 second integration; these three models are readily distinguishable in this data. These model spectra were generated for Sedna's distance from the sun at the time of JWST's launch, and return mean signal-to-noise of 39-42 per raw R1000 spectral element across the three models (maximum signal-to-noise of 76-81 at short wavelengths and minimum signal-to-noise of 1-8 at long wavelengths); these three spectra are formally incompatible with each other at very high confidence.

3. JWST can provide the first concrete detection of complex organics on the surface of small TNOs. Tholins are solids produced in the laboratory when planetary atmosphere analog gases are exposed to a variety of energy sources; they frequently contain organic compounds with both aromatic and aliphatic hydrocarbons (Sagan & Khare, 1979). For many years, tholins have been used as the best analog for modeling the surface of red-colored TNOs and Centaurs, so it has been assumed that organic solids are present on their surfaces, even though diagnostic bands could not be detected (Roush & Cruikshank 2004). Figure 2 contrasts model 10,000 second NIRSpec reflectance spectra of two organics-dominated and one silicate-dominated surfaces for a small (r~80km) Centaur at 22AU, illustrating how readily these three compositions can be distinguished. They return mean signal-to-noise of 16-18 per raw R1000 spectral element across the three models (maximum signal-to-noise of 32-34 at short wavelengths and minimum signal-to-noise of 2-3 at long wavelengths); these three spectra are also formally incompatible with each other at very high confidence.

The detection of aromatic and aliphatic hydrocarbons on Iapetus by Cassini spacecraft (Cruikshank et al. 2005) and on two asteroids 24 Themis (Campins et al. 2010, Rivkin & Emery 2009) and 65 Cybele (Licandro et al. 2011) shows that these materials could be present on the surface of the small bodies in the Solar System. NIRSpec is a promising tool to detect the signature of such complex organics on the surface of red TNOs.

4. Methanol has an absorption band at 2.27 $\mu$m. It has been used to model the spectra of some TNOs (Cruikshank et al. 1998, Merlin et al. 2012). However, the detection using traditional NIR spectrographs from ground based telescopes is very difficult as the band lies on the red extreme of the coverage of these instruments, where the sensitivity is not optimum. Methanol has been suggested as the coloring agent for some red TNOs (Brown et al. 2012). Observations with NIRSpec will provide a test of the models that suggest that methanol is widespread over the trans-Neptunian region and it is responsible for the colors of bright, red objects.

5. For those objects that will not be observable using spectroscopy in a reasonable time, we can rely on photometry. IRAC/Spitzer has been previously used for similar studies (Emery et al. 2007, Dalle-Ore et al. 2009, Pinilla-Alonso et al. 2013). However, IRAC uses two broad band filter to cover from 3 - 6.5 $\mu$m meanwhile JWST offers 7 different filters to cover from 3 - 5 $\mu$m. This more finely-discretized spectral coverage means that the color indices provided by JWST are a much more powerful tool to study the surface compositions of a large population of TNOs. In the NIR, small TNOs can also be spectrally characterized at levels similar to many asteroids; this permits the direct comparison of populations such as the Jupiter Trojans and the Hot Classical KBOs, thought to share a common parent population (Fraser et al. 2014) but have been subjected to different dynamical, collisional, and thermal histories.

Surfaces with compositions similar to the known dwarf planets can be readily distinguished in the IR using NIRCam photometry. Figure 3 illustrates several examples of model NIR reflectance spectra of several dwarf planet surfaces. Additionally, many TNOs (particularly the small Cold Classicals; Peixinho et al. 2008, Benecchi et al. 2011) display highly reddened surfaces; colorants may include silicates and organics. Figure 3 also illustrates that the IR behavior of these materials is drastically different, and the NIRCam filter bandpasses can be used to efficiently identify either material.

Due to NIRCam's sensitivity, a spectrophotometric survey in the IR of many TNOs can be achieved rapidly. For example, a spectrophotometry diagnostic survey similar to H/WTSOSS (Fraser and Brown, 2012) in F182M, F210M, and F300M would be capable of distinguishing silicate-, organics-, or water-dominated surfaces, and could be conducted for nearly every known D>100km TNO within the expected time for TNO science, accounting for known TNO sizes, distances, and target accessibility.

## Community actions

To make the most efficient use of the limited resource of JWST observing time that will be available to the TNO research community, several actions are merited.

**(1) Potential target pre-characterization**

Due to the relatively limited total time likely available to TNO science, it will be critical to accurately measure target properties with existing facilities to improve predictions for JWST performance and identify ideal candidates for expending JWST time. For many TNOs, further astrometric measurements will be required to ensure that the uncertainties in their ephemerides are small enough that JWST can target them. Currently, under 25% of all recoverable (ie., not effectively lost to all current facilities) TNOs have orbits precise enough to be confidently blindly targeted within NIRSpec's 3" x 3" IFU FOV in 2019 without additional astrometric follow-up between now and then. For many currently-untargetable objects, a single additional ground-based astrometric follow up epoch would be sufficient to make NIRSpec targeting feasible.

Ground-based occultation campaigns provide a valuable avenue to both accurately measure diameters (and thus provide critical cross-checks and calibration for JWST radiometric diameters) and to identify objects with unusual properties that merit detailed follow-up with JWST. Concerted efforts to collect high-quality occultation measurements of many Centaurs and TNOs will help to identify objects with rings, atmospheres, and activity (such as the occultation-discovered rings and ring-like structures around Charik-lo and Chiron, Braga-Ribas et al. 2014, Ortiz et al. 2015) which would make ideal targets for characterization with JWST. As the rate of occultation detections is largely limited by the collection rate of new high-quality TNO orbit-refinement astrometry (required in order to accurately predict the timing and location of occultation events), this further supports the need for campaigns of ground- and space-based TNO orbit-refinement astrometry.

Finally, expanding the sample of wide binary systems with well-measured orbital properties will be important for predicting epochs in which the systems will be resolvable with JWST at thermal wavelengths for albedo and diameter measurements; fewer than 10 ultra-wide systems with separations larger than MIRI's ~1" long-wavelength spatial resolution currently have characterized mutual orbits (Parker et al. 2011, Sheppard et al. 2011). Searches for widely-separated binary systems can be achieved with ground-based facilities (all known wide systems were discovered in non-adaptive optics ground-based imagery), and they are natural targets for long-term (several years; Parker et al. 2011) orbital characterization campaigns to measure their mutual orbit properties and derive system masses. While the occurrence rate of these widely-separated systems is currently poorly constrained, any new system would provide a substantial increase in the sample size of candidate TNO binaries resolvable in thermal observations by JWST.

**(2) Expanding laboratory investigation of spectral properties of relevant ices through the optical, NIR and MIR.**

Many chemical species may be present and spectrally active on the surfaces of TNOs accessible to JWST. However, many relevant species lack laboratory characterization of their optical constants over the wavelength range covered by JWST and at temperatures relevant to TNOs. Comprehensive optical constants are lacking for species as ubiquitous as methane and water; $C_2H_6$ and higher hydrocarbons, methanol, ammonia hydrates, many nitriles, phyllosilicates, and salts are also without complete characterization. A concerted effort must be made to expand the library of species and alloys with well-characterized optical constants over TNO temperature ranges covering the wavelength range of JWST.

**Figure 1:** Simulated 10,000 second NIRSpec reflectance spectra of Sedna (D~995 km at 85 AU) for three hypothetical surface compositions. The spectral range of NIRSpec is ideal for disentangling a variety of plausible surface chemistries. The raw R1000 spectra is illustrated with light curves (mean S/N ranging from 39-42 per raw spectral element for the three models; maximum signal-to-noise of 76-81 at short wavelengths and minimum signal-to-noise of 1-8 at long wavelengths), while smoothed versions of the same data is shown with the heavy curves. All three model spectra are formally incompatible with each other at very high confidence.

**Figure 2:** Simulated 10,000 second NIRSpec reflectance spectra of a small Centaur (D~80km at 22AU) for three hypothetical surface compositions. The spectral range of NIRSpec is ideal for disentangling a variety of plausible surface chemistries. The raw R1000 spectra is illustrated with light curves (mean S/N ranging from 16-18 per raw spectral element for the three models; maximum signal-to-noise of 32-34 at short wavelengths and minimum signal-to-noise of 2-3 at long wavelengths), while smoothed versions of the same data is shown with the heavy curves. All three model spectra are formally incompatible with each other at very high confidence.

**Figure 3:** Coverage of NIRCam filter bandpasses compared to model reflectance spectra of a variety of real and hypothetical TNO surface chemistries. **Top left:** Models resembling the surface composition of the dwarf planets whose spectra in the visible and NIR are dominated by methane ice (with Makemake as a template), and those whose spectra are dominated by water ice (with Haumea as a template). **Top right:** Models fitting real objects (Emery et al. 2007, Dalle-Ore et al. 2009). These are red objects (presence of organics), with some water ice in the visible and a minor amount of other constituents. These are the only models published that are derived from a combination of vis+NIR+IRAC/Spitzer data. **Bottom right:** Spectra of two putative reddening agents for surfaces of small TNOs. These two reddening agents are remarkably distinguishable in the IR.

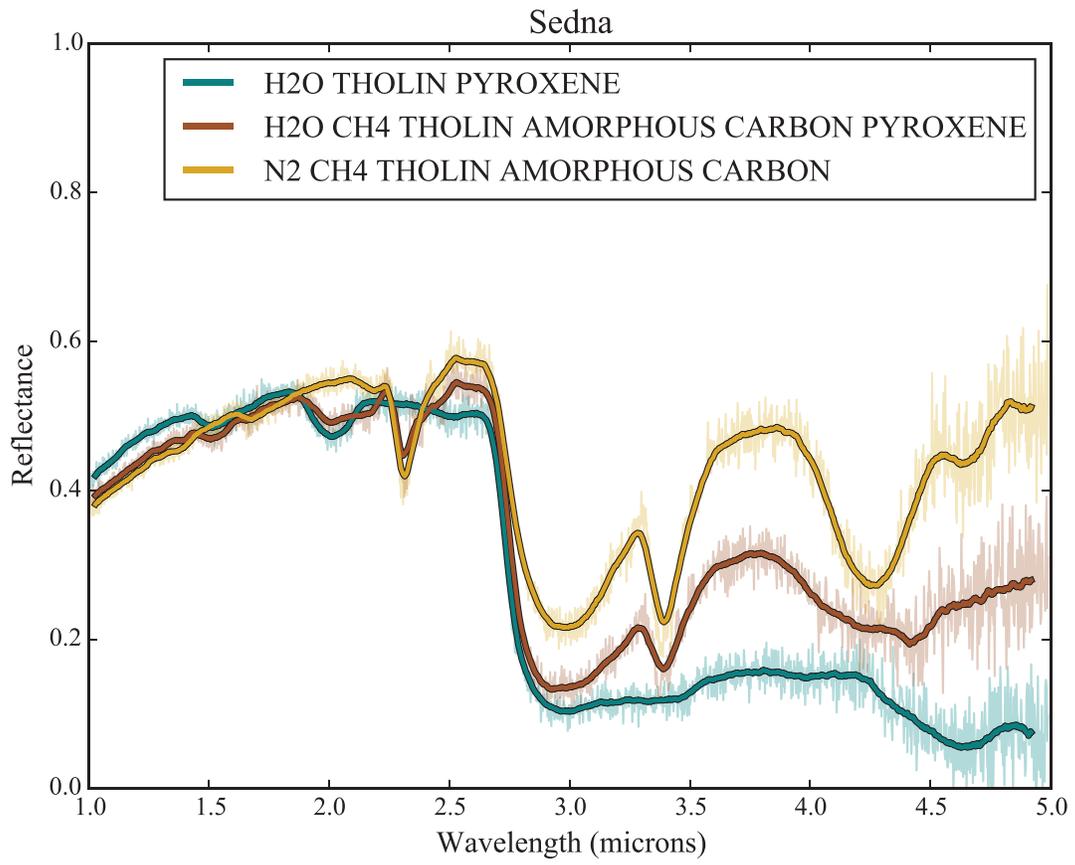

Figure 1

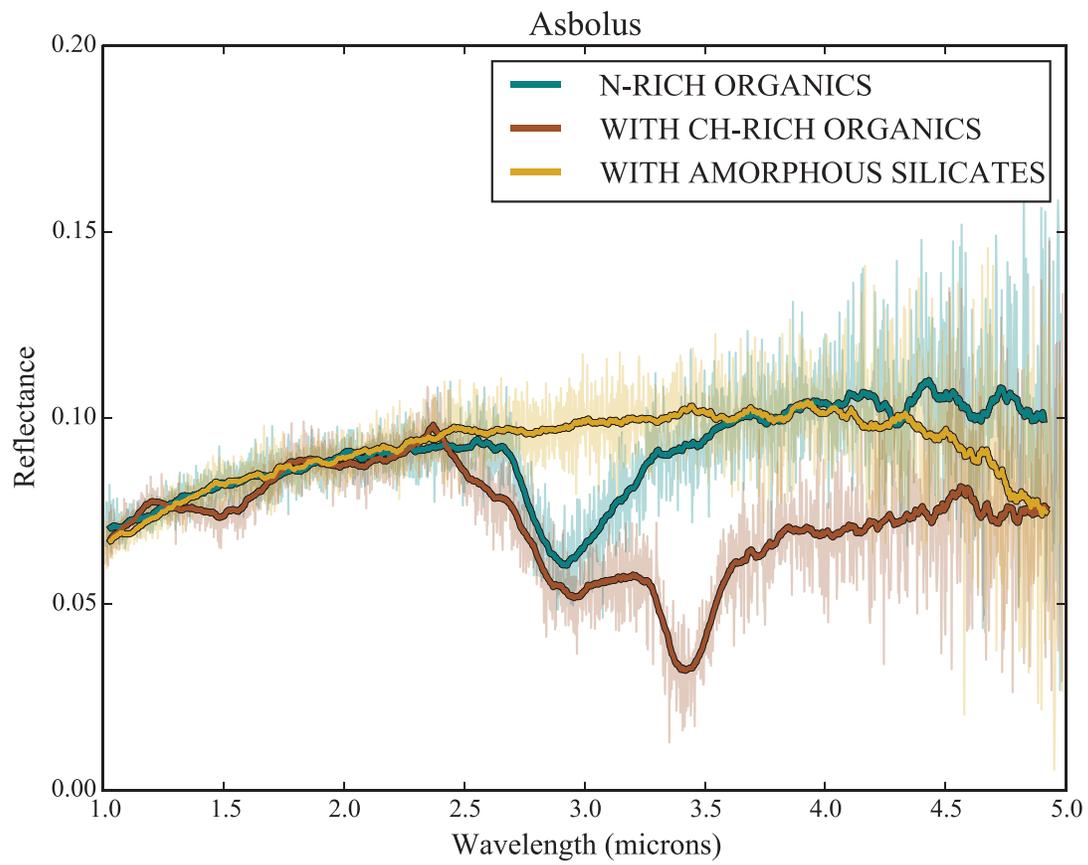

Figure 2

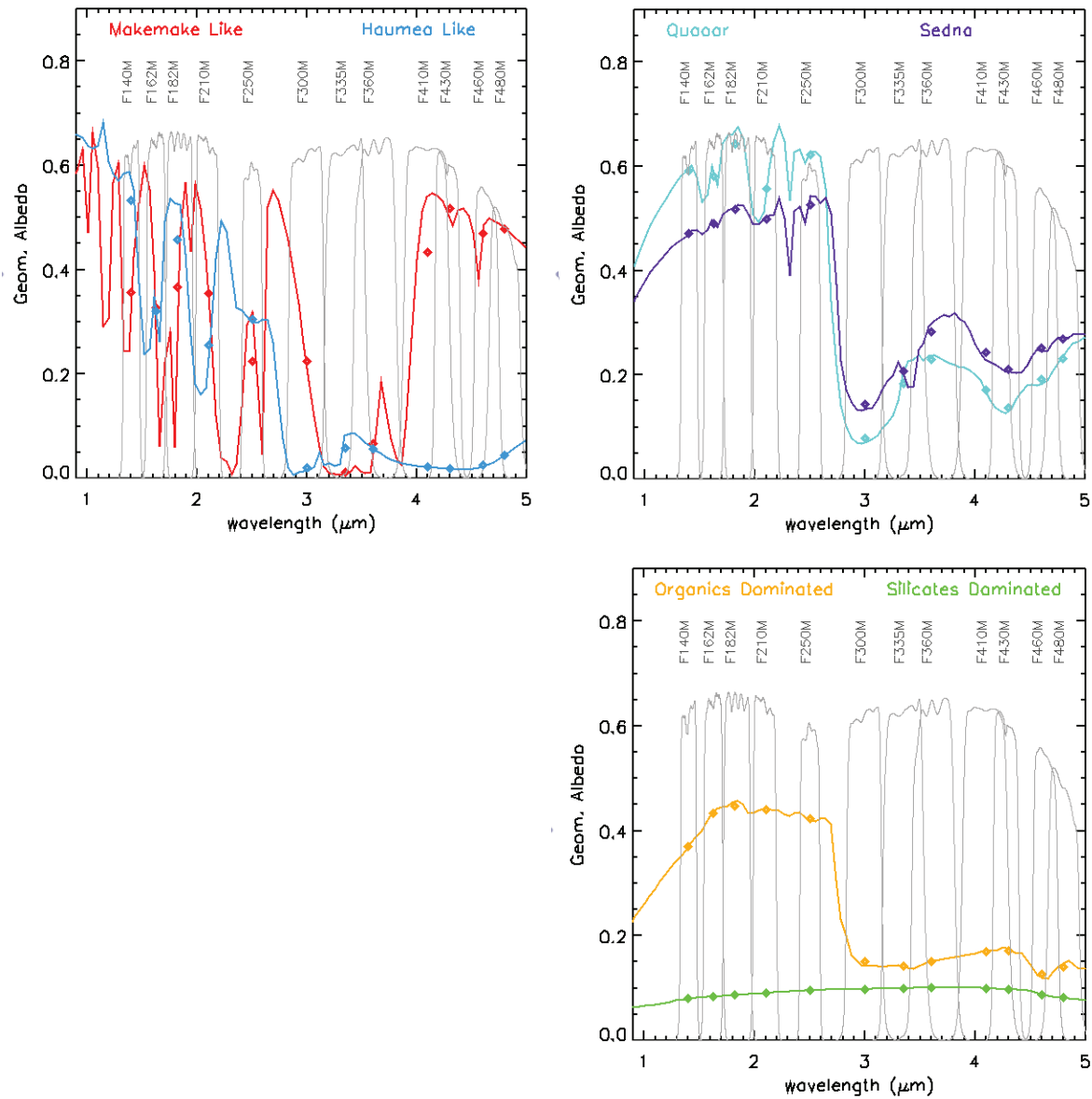

Figure 3